\documentclass[11pt,twocolumn]{article}

\usepackage{times}
\usepackage{a4wide}
\usepackage[final]{graphicx}
\usepackage{amsmath,amssymb}
\usepackage{authblk} % for authors and affiliations
\usepackage[numbers,compress]{natbib}
\usepackage[margin=1em,font={small},labelfont={sc}]{caption}

\usepackage[pagebackref=false,colorlinks=true,citecolor=blue,filecolor=blue,%
linkcolor=blue,anchorcolor=blue,urlcolor=blue,hypertex]{hyperref}

\renewcommand\Re{\operatorname{\mathfrak{Re}}}
\renewcommand\Im{\operatorname{\mathfrak{Im}}}

  \renewenvironment{thebibliography}[1]{%
    \begin{oldthebibliography}{#1}%
      \setlength{\parskip}{0ex}%
      \setlength{\itemsep}{0ex}%
      \small
  }%
  {%
    \end{oldthebibliography}%
  }
\begin{document}

\title{\LARGE \bf Parametrizing Compton form factors with 
neural networks\footnote{Presented by K.K. at
Ringberg Workshop \emph{New Trends in HERA Physics}, 25--28 September 2011.}
}

\author[1]{Kre\v{s}imir Kumeri\v{c}ki}
\author[2,3]{Dieter M\"{u}ller}
\author[4]{Andreas Sch\"{a}fer}
\affil[1]{Department of Physics, University of Zagreb,
              Bijeni\v{c}ka c. 32, 10002 Zagreb, Croatia}
\affil[2]{Brookhaven National Lab, Physics Department, Upton, NY 11973-5000, U.S.}
\affil[3]{Institut f\"ur Theoretische Physik II, Ruhr-Universit\"{a}t Bochum,
              Universit\"{a}tsstra\ss{}e 150, 44780 Bochum, Germany}
\affil[4]{Institut f\"{u}r Theoretische Physik, Universit\"{a}t Regensburg,
              Universit\"{a}tsstra\ss{}e 31, 93053 Regensburg, Germany}
\date{}
\maketitle

\begin{abstract}
\noindent
We describe a method, based on neural networks, of revealing Compton form
factors in the deeply virtual region. We compare this approach to standard
least-squares model fitting both for a simplified toy case and for HERMES data.
\end{abstract}

\section{Introduction}
\label{sec:intro}

Extraction of generalized parton distribution (GPD) functions
\cite{Mueller:1998fv,Radyushkin:1996nd,Ji:1996nm} from
exclusive scattering data is an important endeavour,
related to such practical questions as
the partonic decomposition of the nucleon spin  \cite{Ji:1996ek} and
characterization of multiple-hard reactions in
proton-proton collisions at LHC collider \cite{Diehl:2011tt,Diehl:2011yj}.
To reveal the shape of GPDs, one employs global or local fits to data
\cite{Kumericki:2007sa,Goloskokov:2007nt,Guidal:2008ie,Kumericki:2009uq,Guidal:2009aa,Guidal:2010ig, Moutarde:2009fg}.
However, compared to familiar global parton distribution (PDF) fits, fitting of GPDs is
intricate due to their dependence on three kinematical variables (at
fixed input scale $\mathcal{Q}_0$), and the fact that they
cannot be fully constrained even by ideal data.
Thus, final results can be significantly influenced by the choice of the
particular fitting ansatz.
To deal with this source of theoretical uncertainties, we used
an alternative approach \cite{Kumericki:2011rz}, in which
\emph{neural networks} are used in place of specific models.
This approach has already been successfully applied to extraction of
the deeply inelastic scattering (DIS) structure function $F_2$ and normal PDFs
\cite{Forte:2002fg,Ball:2008by,Ball:2010de}. We expect that the
power of this approach is even larger in the case of GPDs. In the light
of the scarce experimental data, in this pilot study we attempted the
mathematically simpler extraction
of form factor $\mathcal{H}(x_B, t)$ of deeply virtual
Compton scattering (DVCS). We used data from the kinematical region
where this Compton form factor (CFF) dominates the observables and depends
essentially  only on two kinematical
variables: Bjorken's scaling variable $x_B$ and proton momentum transfer squared $t$.
These simplifications make the whole problem more tractable.

\section{The method}
\label{sec:nnmethod}

\begin{figure*}[th]
\centerline{\includegraphics[scale=0.9]{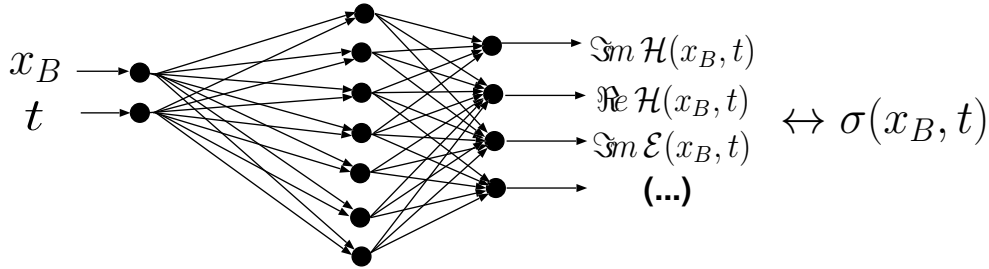}}% two columns
\caption{\small
The structure of a neural network that represents a set of CFFs $\{\mathcal{H}(x_{\rm B}, t), \mathcal{E}(x_{\rm B}, t), \ldots \}$.
The network is trained by calculating observables (cross-sections $\sigma(x_{\rm B}, t)$
or asymmetries) from CFFs, comparing them to experimentally measured values, and then
by adjusting network parameters to minimize the squared errors.
\label{fig:perceptron} }
\end{figure*}

Neural networks were invented some decades ago in an attempt to create
computer algorithms that would be able to classify (i.e.~recognize) complex patterns.
The specific neural network type used in this work, known as \emph{multilayer perceptron},
is a mathematical structure consisting of a number of interconnected
``neurons'' organized in several layers.
It is schematically shown in Fig.~\ref{fig:perceptron}, where each blob
symbolizes a single neuron.
Each neuron has several inputs and one output. The value
at the output is given as a function $f(\sum_j w_j x_j)$ of a sum of input values
$x_1, x_2, \cdots$, each weighted by a certain number
$w_j$.
%For \emph{activation function} $f(x)$ we employed nonlinear logistic sigmoid function
%f(x) = 1/(1+\exp(-x))$
%or neurons in inner (``hidden'') layer, while for input and output layers we
%sed the identity function $f(x)=x$.

The parameters of a neural network (weights $w_j$) are adjusted by a procedure known
as ``training'' or ``learning''. Thereby, the input part of a chosen set of training
input-output patterns is presented to the
input layer and propagated through the network to the output layer.
The output values are then compared to known values of the output part of training
patterns and the calculated differences are used to adjust the network weights.
This procedure is repeated until the network can correctly
classify all (or most of all) input patterns. If this is done properly,
the trained neural network is capable
of generalization, i.e., it can successfully classify patterns it
has never seen before.

This whole paradigm can be applied also to fitting of functions to
data. Here, measured data are the patterns, 
the input are the values of the kinematical                                                      
variables the observable in question depends upon,
and the output is the value of this observable, 
see Fig.~\ref{fig:perceptron}.
%In classification problems, a neural
%network represents a map from a pattern
%space (sometimes of enormous dimensionality) into a simple discrete set of classes,
%whereas in fitting problems neural network represents a map from
%simpler space (two-dimensional $x_B$--$t$ space in the case of Fig.~\ref{fig:perceptron})
%into a space of one or several real-valued functions.
In this case, the generalization property of neural networks
represents its ability to provide a reasonable estimate of the actual underlying
physical law. For the particular application of neural networks to fits of
hadron structure functions we refer the reader to papers of the NNPDF
group \cite{Forte:2002fg,Ball:2008by,Ball:2010de,Rojo:2006ce}. Our approach is
similar and is described in detail in \cite{Kumericki:2011rz,Kumericki:2011zt}.

To propagate experimental uncertainties into the final result, we use the
``Monte Carlo'' method \cite{Giele:2001mr}, where neural networks are not
trained on actual data but on a collection of ``replica data sets''. These sets
are obtained from original data by generating random artificial data points
according to Gaussian probability distribution with a width defined by the
error bar of experimental measurements.  Taking a large number $N_{rep}$ of
such replicas, the resulting collection of trained neural networks
$\mathcal{H}^{(1)},\ldots,\mathcal{H}^{(N_{rep})}$
defines a probability distribution $\mathcal{P}[\mathcal{H}]$ of the represented CFF
$\mathcal{H}(x_B, t)$ and of any functional $\mathcal{F}[\mathcal{H}]$
thereof. Thus, the mean value of such a functional and
its variance are \cite{Giele:2001mr,Forte:2002fg}
\begin{align}
 \Big\langle \mathcal{F}[\mathcal{H}] \Big\rangle& =
 \int \mathcal{D}\mathcal{H}
 \: \mathcal{P}[\mathcal{H}] \, \mathcal{F}[\mathcal{H}]  \nonumber \\
 & =
  \frac{1}{N_{rep}}\sum_{k=1}^{N_{rep}} \mathcal{F}[\mathcal{H}^{(k)}]\;,
\label{eq:funcprob} \\
\Big(\Delta \mathcal{F}[\mathcal{H}]\Big)^2& =
\Big\langle \mathcal{F}[\mathcal{H}]^2 \Big\rangle  -
\Big\langle \mathcal{F}[\mathcal{H}] \Big\rangle^2 \;.
\label{eq:variance}
\end{align}

\section{Toy example}
\label{sec:toy}

\begin{figure*}[th]
\begin{center}
\includegraphics[scale=0.44]{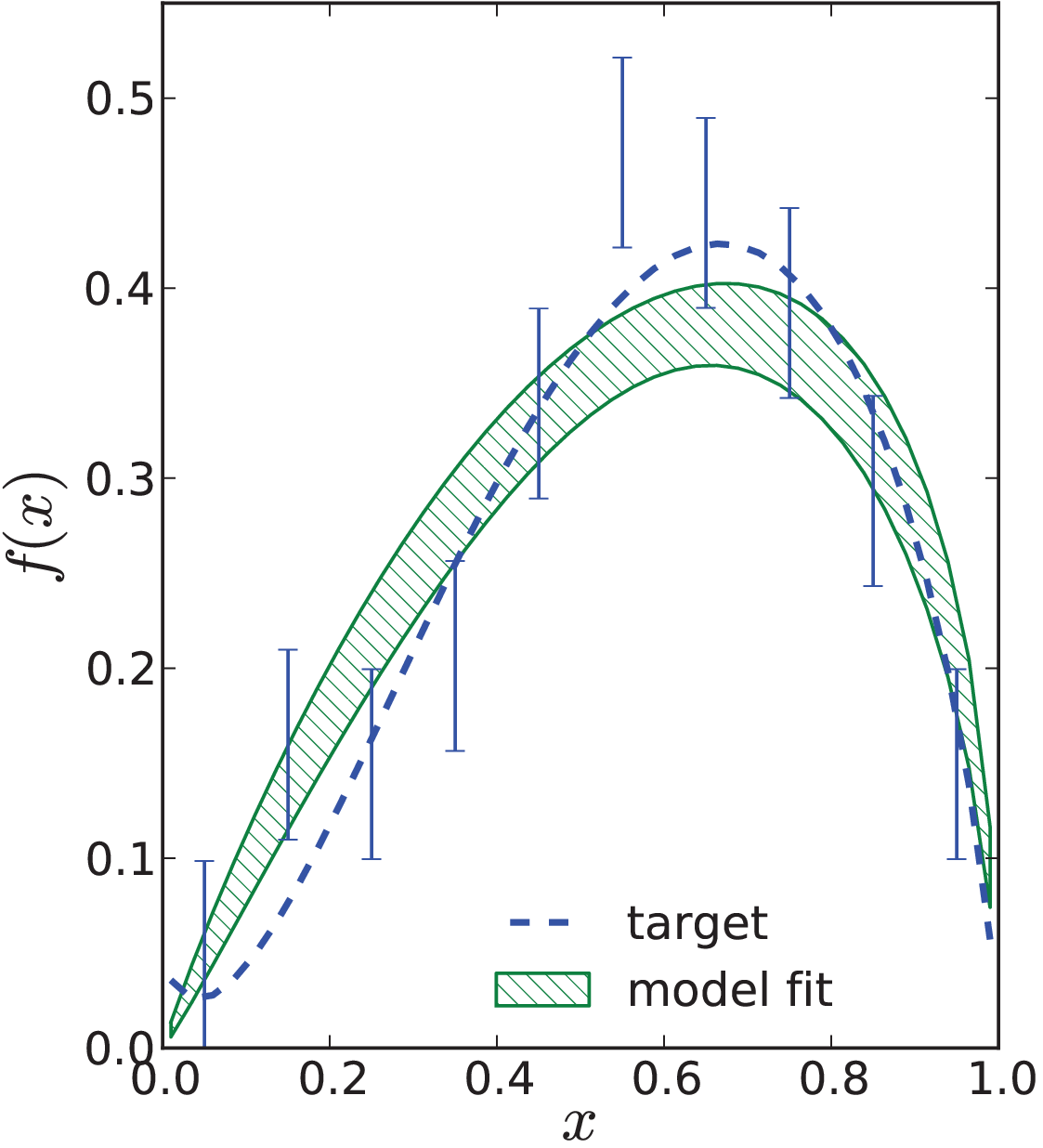}%
\includegraphics[scale=0.44]{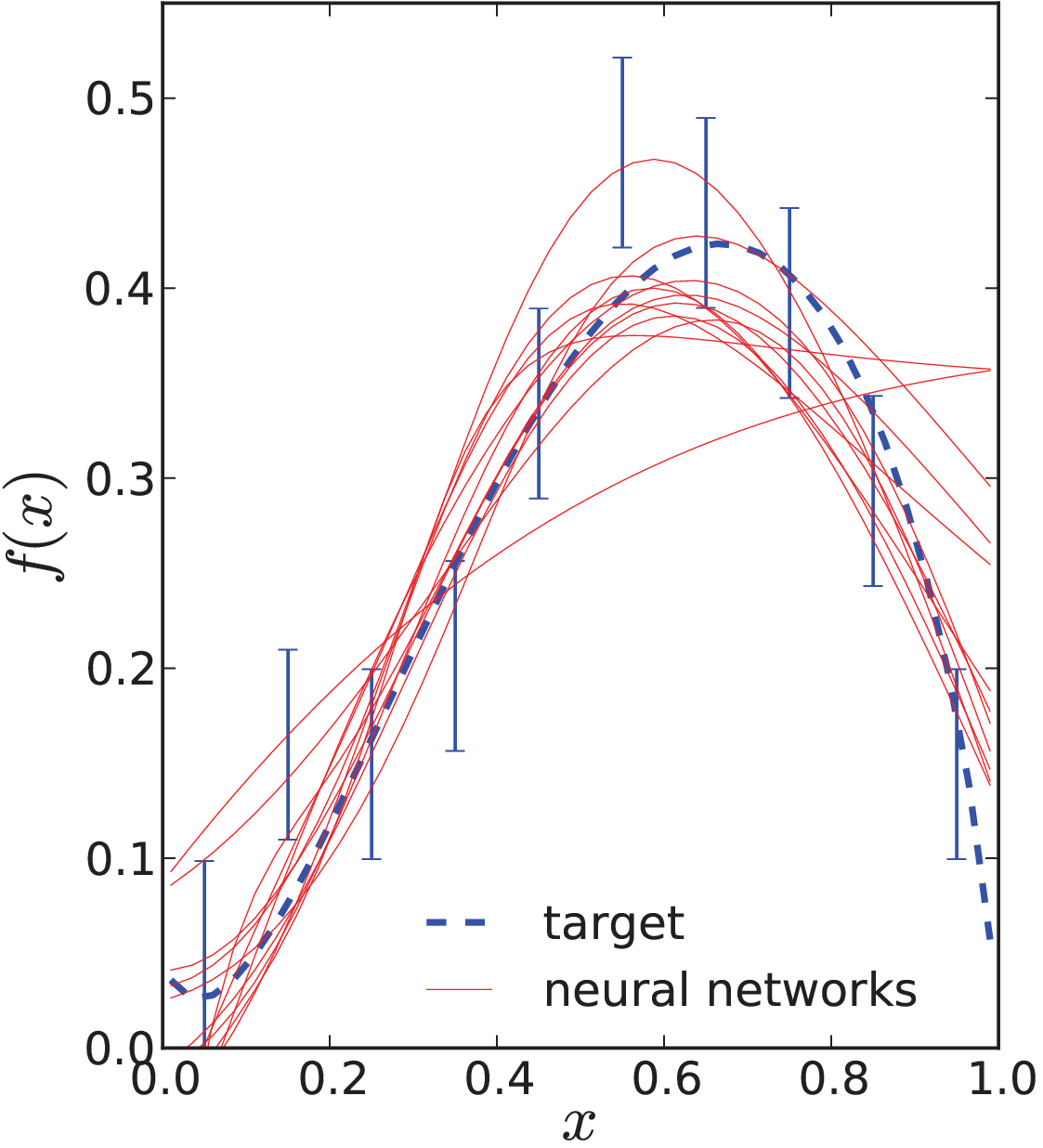}%
\includegraphics[scale=0.44]{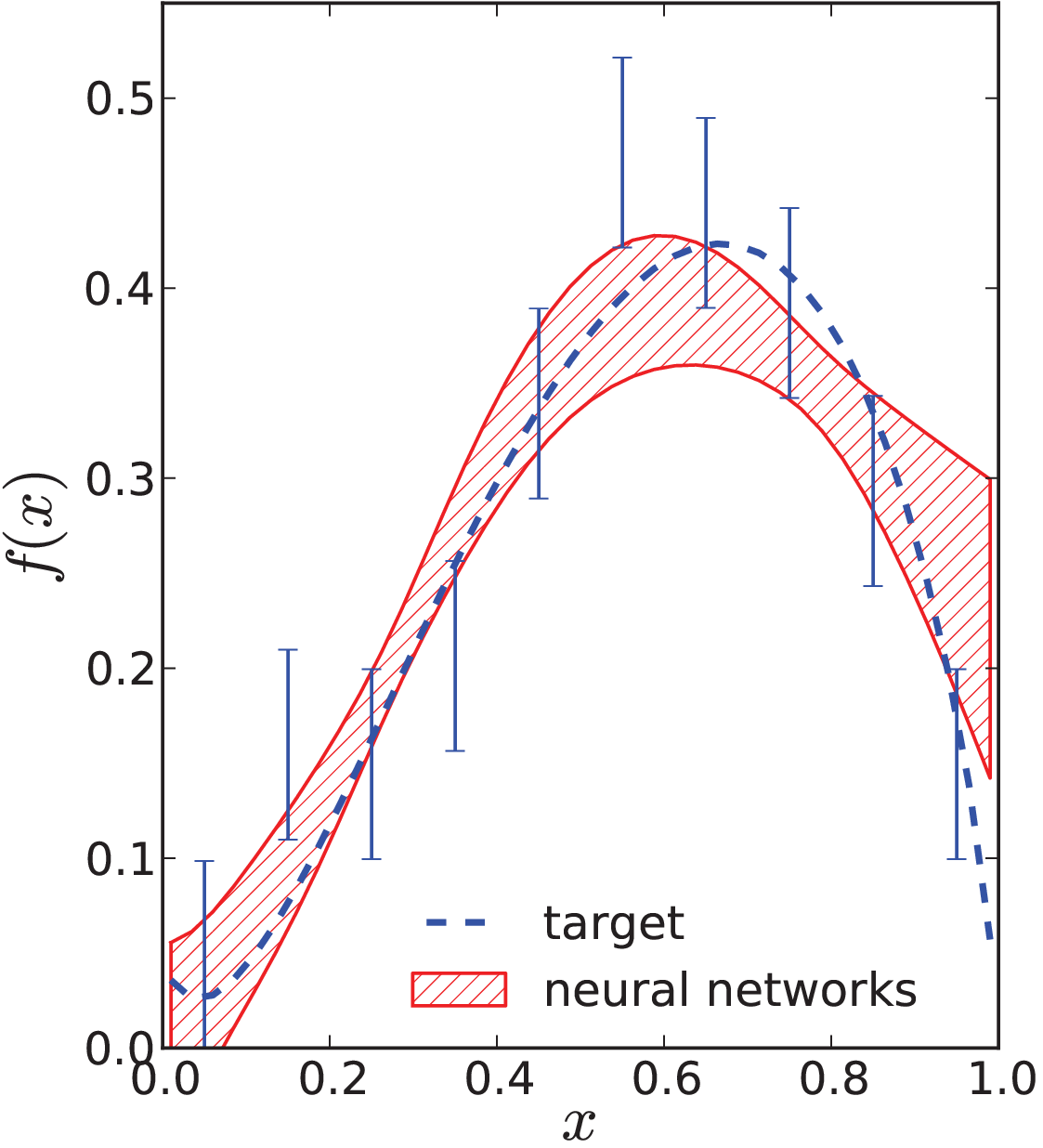}
\end{center}
\caption{Toy examples of fitting to fake data, generated from the underlying target function (dashed).
The first panel shows the result of a standard least-squares model fit, the second one shows twelve neural networks that are trained
on the Monte Carlo replicas of fake data, and the third panel shows the uncertainty band obtained
by statistical averaging of neural networks (displayed in the second panel).}
\label{fig:toy}
\end{figure*}

To illustrate the neural network fitting method, we shall
now present a toy example where we will extract a known function of
one variable by fitting to fake data. First we define some simple target function $\tilde{f}(x)$
as a random composition of simple polynomial and logarithm functions constrained
by the property
\begin{equation}
    \tilde{f}(1) = 0 \;.
\label{eq:endpoint}
\end{equation}
This function is plotted in Fig.~\ref{fig:toy} as a thick dashed line and
labeled as ``target''.

Next, $n_{\rm pts}$=10 fake data points $(x_i, y_i \pm \Delta y_i)$ are generated
equidistantly in $x$. Their mean
values $y_{i}$ are smeared around  target values by random Gaussian fluctuations
with standard deviation  $\Delta y_i$=0.05, which is also taken to be the uncertainty of
generated points.
These fake data are then used for fits, first using the standard
least-squares method with a two-parameter model
\begin{equation}
 f(x) = x^{p_1} (1-x)^{p_2} \;,
\label{eq:ansatz}
\end{equation}
and, second,  utilizing the neural network method.
Note that the Monte Carlo method of error propagation, which we use together with
neural network fitting, itself requires to generate artificial data sets.
Thus, we generated $N_{rep}$=12 replicas from original fake data and used them to
train 12 neural networks that represent 12 functions, plotted as thin
solid lines on the second panel of Fig.~\ref{fig:toy}.
These functions define a probability distribution in the space of
functions $f(x)$ which, according to Eqs.~(\ref{eq:funcprob}--\ref{eq:variance}),
provides an estimate of the sought function $\tilde{f}(x)$, together
with its uncertainty. This estimate is shown on the right panel of Fig.~\ref{fig:toy} as a
(red) band with ascending hatches. The corresponding model fit result, obtained
by the standard method of least-squares optimization and error propagation using
the Hessian matrix, is shown in the left panel of Fig.~\ref{fig:toy} as a (green) band with
descending hatches.

We have deliberately chosen the ansatz (\ref{eq:ansatz}) with two
properties, incorporating theoretical biases about endpoints: $f(1) = 0$ and $f(0) = 0$.
The first of these actually ``corresponds to the truth'', i.e., to Eq.~(\ref{eq:endpoint}),
whereas the second one is erroneous. As a result, for $x\to 1$ the model fit is in much better
agreement with the target function (thick dashed line) than neural networks, which rely
only on data and are insensitive to this endpoint behaviour.
On the other side, for $x\to 0$ the  model fit is in some small disagreement with the target function,
and, what is much worse, it very much underestimates the uncertainty of the fitted function there
(the uncertainty becomes zero at endpoints!), demonstrating the dangers of unwarranted theoretical
prejudices.

We can be more quantitative and say that according to the standard $\chi^2$ measure,
\begin{displaymath}
 {\chi}^{2} \equiv \sum_{i}^{n_{\rm pts}} \frac{ (y_i - f(x_i))^2}{\Delta y_i^2} \;,
\end{displaymath}
both methods lead to functions that correctly describe data\footnote{We ignore here
the difference between the number of data points $n_{\rm pts}$ and the degrees of freedom --- neural
networks have very many free parameters and for them degrees of freedom is not such an important characteristic
as in the case of standard model fits.}:
\begin{align*}
\chi^{2}_{\rm model}/n_{\rm pts}& = 11.9/10 \;; \\
\chi^{2}_{\rm neur. net}/n_{\rm pts}& = 12.3/10 \;.
\end{align*}
We can now further ask to what extent the two methods extract the underlying target
function $\tilde{f}(x)$. Naturally, we can measure this by a kind of $\bar{\chi}^{2}$ criterion
\begin{displaymath}
 \bar{\chi}^{2} \equiv \sum_{i}^{n_{\rm pts}} \frac{ (\tilde{f}(x_i) - f(x_i))^2}{\Delta f(x_i)^2} \;,
\end{displaymath}
where the denominator is now the propagated uncertainty $\Delta f(x_i)$  rather than the
experimental one $\Delta y_i$. In our toy example we get
\begin{align*}
\bar{\chi}^{2}_{\rm model}/n_{\rm pts}& = 25.6/10 \;; \\
\bar{\chi}^{2}_{\rm neur. net}/n_{\rm pts}& = 8.4/10 \;,
\end{align*}
showing that the model fit underestimates its uncertainties, while neural networks
are much more realistic.

\begin{figure*}[t]
\centerline{\includegraphics[scale=0.55,clip]{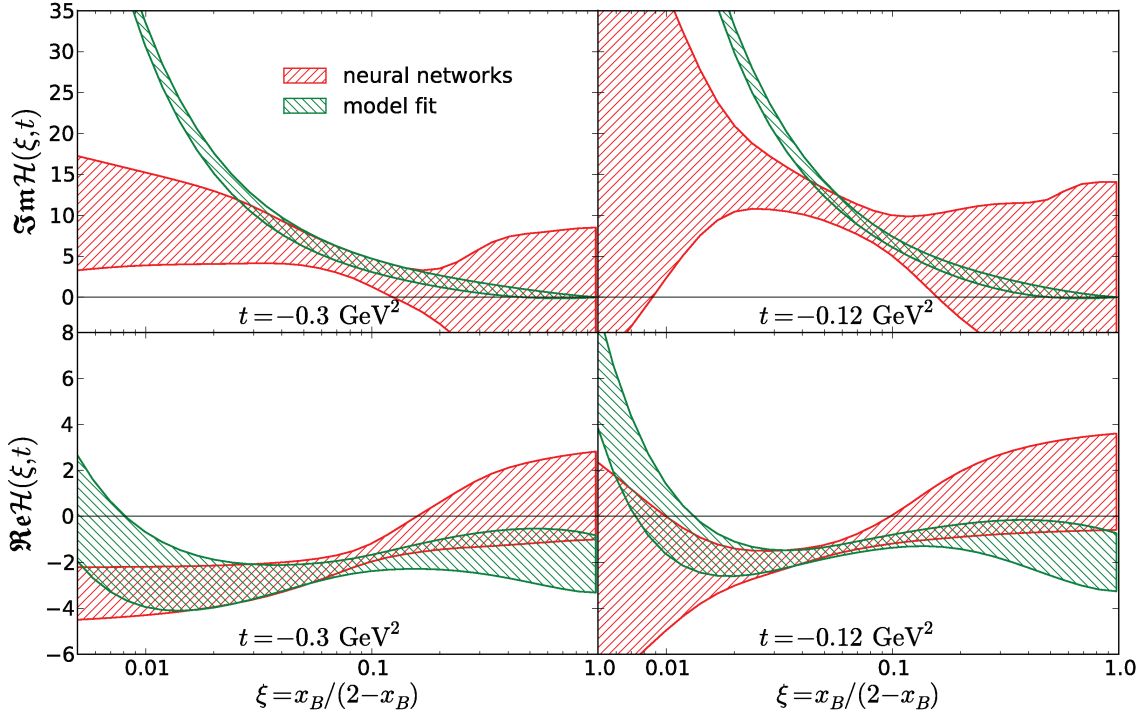}}%
\caption{\small
Neural network extraction of $\Im{\cal H}(x_{\rm Bj},t)$
and $\Re{\cal H}(x_{\rm Bj},t)$ (ascending hatches, red)
from HERMES data~\cite{:2009rj} compared with model
fits (descending hatches, green) for two different values
of momentum transfer squared $t$.
\label{fig:cff} }
\end{figure*}

This example shows that the neural network method has a clear advantage if we want
bias-free propagation of information from experimental measurements into the CFFs.
Still, if we want to use  some additional input,
e.g., if we rely on the spectral property (\ref{eq:endpoint}), 
we can do so also within the neural network method.
For example, we could take the output of neural networks
in this toy example not as an representation of the function $f(x)$ itself, but as representing
$f(x)/(1-x)^{p}$, with some positive power $p$. Then the final neural network predictions for $f(x)$
would also be constrained by Eq.~(\ref{eq:endpoint}), without any further loss of generality (in practice
it turns out that the dependence of the results on the choice of power $p$ is small).
Various methods of implementing theoretical constraints in the neural network
fitting method are discussed in Sect.~5.2.4 of \cite{Rojo:2006ce}.

\section{Application to HERMES data}
\label{sec:fit}

To extract the CFF $\mathcal{H}$  from asymmetries \cite{:2009rj}, measured by the HERMES  collaboration in photon electroproduction off unpolarized protons, we applied the described neural network fitting method in \cite{Kumericki:2011rz}.  We used 36 data
points: 18 measurements of the first sine harmonic
$A_{LU}^{\sin\phi}$ of the beam spin asymmetry, and 18 measurements of
the first cosine harmonic $A_{C}^{\cos\phi}$ of the beam charge asymmetry. 
As for the toy model from the previous section, we compare the results with
the standard least-squares model fit. Let us first shortly describe this
model fit of $\mathcal{H}$.
For the partonic decomposition of the imaginary part $\Im\mathcal{H}$
we used %a version of 
a model, presented in \cite{Kumericki:2009uq}:
\begin{displaymath}
\Im{\cal H}(x_{\rm Bj},t) = \pi \left[
H^{\rm val}(\xi,\xi,t) + \frac{2}{9} H^{\rm sea}(\xi,\xi,t) \right] .
\end{displaymath}
Here, $H^{\rm a}(\xi, \xi, t)$ are  GPDs along the cross-over trajectory $\xi=x$,
parameterized as:
\begin{multline*}
\label{GPD-Ans}
%\hspace*{-4ex}
H(x,x,t)  =
\frac{n\, r}{1+x} \left(\frac{2 x}{1+x}\right)^{-\alpha(t)} 
\\   \times \left(\frac{1-x}{1+x}\right)^{b}
\frac{1}{\left(1-  \frac{1-x}{1+x} \frac{t}{M^2}\right)^{p}}\;.
\end{multline*}
The parameters of $H^{\rm sea}$ were fixed by 
separate fits \cite{Kumericki:2009uq} to collider data,
and some parameters of $H^{\rm val}$ were also fixed using information from
DIS data and Regge trajectories $\alpha(t)$. The real part $\Re\mathcal{H}$
is expressed in terms of the imaginary one via a dispersion integral
\cite{Teryaev:2005uj,Kumericki:2007sa,Diehl:2007jb,Kumericki:2008di} and
the subtraction constant $C$, leaving us finally with a model that possesses
four parameters: $r^{\rm val}$, $b^{\rm val}$, $M^{\rm val}$ and $C$.
This model is fitted to experimental data, resulting in parameter values, which can
be found in \cite{Kumericki:2011rz}, and shapes of
$\Im\mathcal{H}$ and $\Re\mathcal{H}$ that are plotted on Fig.~\ref{fig:cff}
as (green) bands with descending hatches.

The neural network fit was performed by creating 50 neural networks with two neurons in the input layer
(corresponding to kinematical variables $x_B$ and $t$), 13 neurons
in the hidden middle layer, and two neurons in the output layer
(corresponding to $\Im\mathcal{H}$ and $\Re\mathcal{H}$), cf.~Fig.~\ref{fig:perceptron}.
These were trained on $N_{rep}$=50 Monte Carlo replicas of HERMES data. We
checked that the resulting CFF $\cal H$
does not depend significantly on the precise number of neurons in the
hidden layer.
The results are also presented on Fig.~\ref{fig:cff}, where
we show the neural network representation
of $\Im\mathcal{H}$ and $\Re\mathcal{H}$ as (red) bands with ascending hatches.

Comparing the two approaches, one notices that in the kinematic region of
experimental data (roughly the middle-$x_B$ parts of Fig.~\ref{fig:cff} panels)
neural network and model fit results coincide, i.e., error bands
are of similar width and they overlap consistently. However, outside of this data region,
we see that the predictions of the two approaches can be different.
There the uncertainty of the model fit is in general smaller, and
we observe a strong disagreement in the low $x_B$ region, reflecting the
theoretical bias of the chosen model that possesses a  $x^{-\alpha(t)}$ Regge
behavior. The lesson learned from the toy model example is that, even if
we believe in Regge behaviour for small $x_B$, we should still consider the
uncertainty from the neural network method as more realistic.

\section{Conclusion}

Utilizing both a simplified toy example and HERMES measurements of photon electroproduction  asymmetries, 
we demonstrated that neural networks and Monte Carlo error propagation provide
a powerful and unbiased tool that extracts information
from data. Comparisons with standard least-squares model fits reveal that the uncertainties, obtained from neural network fits, 
are reliable and realistic.

Relying on the hypothesis of $\mathcal{H}$ dominance, we found the  CFF $\mathcal{H}$ from a completely unconstrained neural network fit.
It is expected that the extraction of all four leading twist-two CFFs 
($\mathcal{H}$, $\mathcal{E}$, $\widetilde{\mathcal{H}}$ and
$\widetilde{\mathcal{E}}$, or the corresponding GPDs)
from presently or soon-to-be available data will still be an ill-defined optimization problem. 
Thus, it might be necessary to implement in neural network fits some carefully
chosen theoretically robust constraints,
such as dispersion relations, sum rules \cite{Kumericki:2008di} and lattice input.



%% References

\section*{Acknowledgments}

This work was supported by the BMBF grant under the contract no. 06RY9191,
by EU FP7 grant HadronPhysics2, by DFG grant, contract no. 436 KRO 113/11/0-1 and by
Croatian Ministry of Science, Education and Sport, contract no.
119-0982930-1016.

%\nocite{*}
%\bibliographystyle{elsarticle-num}  % This is producing too much stuff !!
% Closest to what they would probably like to have is:
%\bibliographystyle{h-physrev4.bst}
%\bibliography{/home/kkumer/Lit/kkumer}

% if you insist on using elsarticle-num.bst, comment out the line above and uncomment the following one
%\input{kumericki-elsarticle-num.bbl}
\end{document}